\documentclass[aps,prl,twocolumn,superscriptaddress,showpacs]{revtex4}
\usepackage{graphicx}
\usepackage{mathrsfs}
\usepackage{bm}

\begin{document}

\title{ Strongly gravitational lensed SNe Ia as multi-messengers: Direct test of the Friedman-Lema\^{\i}tre-Robertson-Walker metric}
\author{Jingzhao Qi}
\affiliation{School of Physics and Technology, Wuhan University,
Wuhan 430072, China;} \affiliation{Department of Astronomy, Beijing
Normal University, Beijing 100875, China;} \affiliation{Department
of Physics, College of Sciences, Northeastern University, Shenyang
110819, China;}
\author{Shuo Cao \footnote{caoshuo@bnu.edu.cn}}
\affiliation{Department of Astronomy, Beijing Normal University,
Beijing 100875, China;}
\author{Marek Biesiada}
\affiliation{Department of Astronomy, Beijing Normal University,
Beijing 100875, China;} \affiliation{Department of Astrophysics and
Cosmology, Institute of Physics, University of Silesia, 75 Pu{\l}ku
Piechoty 1, 41-500, Chorz{\'o}w, Poland}
\author{Xiaogang Zheng}
\affiliation{School of Physics and Technology, Wuhan University,
Wuhan 430072, China;}
\author{Xuheng Ding}
\affiliation{School of Physics and Technology, Wuhan University,
Wuhan 430072, China;}
\author{Zong-Hong Zhu \footnote{zhuzh@bnu.edu.cn}}
\affiliation{School of Physics and Technology, Wuhan University,
Wuhan 430072, China;} \affiliation{Department of Astronomy, Beijing
Normal University, Beijing 100875, China;}

\begin{abstract}

We present a new idea of testing the validity of the
Friedman-Lema\^{\i}tre-Robertson-Walker metric, through the multiple
measurements of galactic-scale strong gravitational lensing systems
with type Ia supernovae in the role of sources. Each individual
lensing system will provide a model-independent measurement of the
spatial curvature parameter referring only to geometrical optics
independently of the matter content of the universe. This will
create a valuable opportunity to test the FLRW metric directly. Our
results show that with hundreds of strongly lensed SNe Ia observed
by LSST, one would produce robust constraints on the spatial
curvature with accuracy $\Delta \Omega_k=0.04$ comparable to the
Planck 2015 results.

\end{abstract}


\maketitle

\textit{Introduction.---}  Friedman-Lema\^{\i}tre-Robertson-Walker
(FLRW) metric is based on the homogeneity and isotropy of the
Universe, which is supported by observations of the large-scale
distribution of galaxies and the near-uniformity of the CMB
temperature \citep{Ade16}. Moreover, it provides the context for
interpreting the observed cosmic acceleration, one of the most
important issues of modern cosmology \citep{Riess98,Perlmutter99}.
However, departure from the FLRW approximation could potentially
explain the late-time cosmic acceleration \citep{Boehm13,Redlich14},
while growing observational data of increased precision enabled
testing the robustness of the FLRW metric
\cite{Clarkson08,Shafieloo10,Mortsell11,Sapone14,Rasanen14}. In
particular, it was proposed that strong lensing data could provide a
consistency test of the cosmic curvature
\citep{Rasanen15,Liao17a,Xia17,Qi19}. However, this method makes a
strong assumption based on the isotropy and homogeneity of the
Universe, that \textit{the distance indicators (SNe Ia, etc.) should
provide the distance information exactly applicable to
galactic-scale strong lensing systems at the same redshift}.

In this letter, we propose a new idea of testing the FLRW metric,
through the multiple measurements of galactic-scale strong lensing
systems with SNe Ia as background sources \citep{Cao18}. Strongly
gravitationally lensed SNe Ia (SGLSNe Ia thereafter), which have
long been predicted in the literature long ago
\citep{Refsdal64,Oguri10}, had not been discovered until very
recently \cite{Goobar17,More17}. The advantage of our method is
that, I) it is independent of the matter content of the Universe and
its relation to space-time geometry; II) each individual lensing
system provides a cosmological model-independent measurement,
without any redshift correspondence from other observations.
Therefore, with a sample of measurements of cosmic curvature at
different sky positions, one could directly test the validity of the
FLRW metric, which would be ruled out if the sum rule was violated
for any pair of lensing systems. Moreover, if the sum rule was
consistent with observations, this test would provide a measurement
of the spatial curvature of the Universe.

\textit{Method.---} Strong gravitational lensing occurs whenever the
source, lens and observer are well aligned that the observer-source
direction lies inside the so-called Einstein radius of the lens. We
will focus on gravitational lensing caused by a galaxy-sized lens.
For a SGL system with the lensing galaxy (at redshift $z_l$),
angular separation of multiple images of the source (at redshift
$z_s$) depends on the ratio of angular-diameter distances between
lens and source $D^{A}_{l,s}$ and between observer and source
$D^{A}_{s}$. Introducing dimensionless comoving distances
$d_{ls}\equiv d(z_l,z_s)$, $d_l\equiv d(0,z_l)$ and $d_s\equiv
d(0,z_s)$, the distance-sum-rule reads \citep{Rasanen15}
\begin{equation} \label{smr}
\frac{d_{ls}}{d_s}=\sqrt{1+\Omega_kd_l^2}-\frac{d_l}{d_s}\sqrt{1+\Omega_k
d_s^2}.
\end{equation}
Therefore, $\Omega_k$ could be directly derived from the distance
ratio $d_{ls}/d_s$, provided the other two distances are known.

I. The angular diameter distance ratio can robustly be determined by
measurement of the Einstein radius
\begin{equation} \label{Eradius}
 \theta_E =
 \left(\frac{4GM_E}{c^2}\frac{D^{A}_{ls}}{D^{A}_{l}D^{A}_{s}}\right)^{1/2}
\end{equation}
where $M_E$ is the mass enclosed in the cylinder of radius equal to
$\theta_E$. We assume the spherically symmetric power-law mass
distribution $\rho \sim r^{- \gamma}$, commonly used in studies of
lensing caused by early-type galaxies \citep{Treu06,Li16,Ma19}.
After solving the spherical Jeans equation \citep{Koopmans05},
assuming that stellar and total mass distributions follow the same
power-law and velocity anisotropy vanishes, one obtains
\begin{equation} \label{Einstein}  \frac{d_{ls}}{d_s} = \frac{D^{A}_{ls}}{D^{A}_s} =   \frac{\theta_E} {4 \pi}
\frac{c^2}{\sigma_{ap}^2}  \left( \frac{\theta_E}{\theta_{ap}}
\right)^{\gamma-2} f(\gamma, M_E)^{-1}
\end{equation}
where $f(\gamma, M_E)$ is a function of the radial mass profile
slope and $\sigma_{ap}$ is the luminosity averaged line-of-sight
velocity dispersion inside the aperture $\theta_{ap}$
\cite{Cao12,Cao15}.

II. Light rays from multiple images of the lensed source need
different time to complete their travel along different paths and
experience different Shapiro delays. Accurate observations of
photometric light curves of the SNe Ia images
$\boldsymbol{\theta}_i$ and $\boldsymbol{\theta}_j$ will provide
time delays \citep{Courbin11}, which are directly related to the
lens potential as well as the mutual distances in the lensing system
\cite{Treu10}
\begin{equation}
\Delta t_{i,j} = \frac{D_{\mathrm{\Delta
t}}(1+z_{\mathrm{l}})}{c}\Delta \phi_{i,j}, \label{relation}
\end{equation}
where the Fermat potential difference
$\Delta\phi_{i,j}=[(\boldsymbol{\theta}_i-\boldsymbol{\beta})^2/2-\psi(\boldsymbol{\theta}_i)-(\boldsymbol{\theta}_j-\boldsymbol{\beta})^2/2+\psi(\boldsymbol{\theta}_j)]$
depends on the source position $\boldsymbol{\beta}$ and the
two-dimensional lensing potential $\psi$ satisfies the corresponding
Poisson Equation: $\nabla^2\psi=2\kappa$, where $\kappa$ is the
surface mass density of the deflector in units of the critical
density. The so-called time-delay distance introduced in
Eq.~(\ref{relation}) can be expressed as
\begin{equation}
D_{\mathrm{\Delta t}}\equiv\frac{D^A_{\mathrm{l}}
D^A_{\mathrm{s}}}{D^A_{\mathrm{ls}}}=\frac{c}{1+z_l}\frac{\Delta t_{i,j}}{\Delta
\phi_{i,j}},
\end{equation}
From measurements of $\Delta t_{i,j}$ and $\Delta \phi_{i,j}$
providing the time-delay distance, combined with the distance ratio
$D^{A}_{ls}/D^{A}_s$, one gets the distance
\begin{equation}
D_l=(1+z_l)D_{\mathrm{\Delta t}} \frac{D^{A}_{ls}}{D^{A}_s}.
\end{equation}

III. SNe Ia can be calibrated as standard candles providing
luminosity distances $D^L_s$ through their distance moduli
$\mu_D=m_X-M_B-K_{BX}$ \citep{Qi18}, where $m_X$ is the peak
apparent magnitude in the filter $X$, $M_B$ is its rest-frame B-band
absolute magnitude, and $K_{BX}$ denotes the cross-filter
K-correction \citep{Kim96}. In our context, the unlensed SN Ia flux
should be scaled up by a magnification factor $\mu$ due to
gravitational lensing appropriately. Therefore, the comoving
distance from the observer to the source is
\begin{equation}
D_s=\frac{10^{(\mu_D+2.5log\mu)/5-5}}{1+z_s} (Mpc).
\end{equation}

Now one is able to determine
\begin{equation} \label{k}
\Omega_k (z_l, z_s) = \frac{d_l^4 + d_s^4 + d_{ls}^4 - 2 d_l^2 d_s^2
- 2 d_l^2 d_{ls}^2 - 2 d_s^2 d_{ls}^2}{4 d_l^2 d_s^2 d_{ls}^2}
\end{equation}
in which the dimensionless comoving distances $d$ are related to the
comoving distances $D$ as $D=cd/H_0$. This function is general, but
in the FLRW space-time it should be equal to the present value of
the spatial curvature parameter $\Omega_{k,0}$ and thus should give
the same result for any pair of source and lens. Due to Strong
covariance between $d_l$, $d_s$, and $d_{ls}$, instead of
propagating distance uncertainties, we use Monte-Carlo simulation to
project uncertainties in the lens mass profile, time delays, Fermat
potential difference, and the magnification effect onto the final
uncertainty of $\Omega_k (z_l, z_s)$.

\begin{figure}
\includegraphics[scale=0.3]{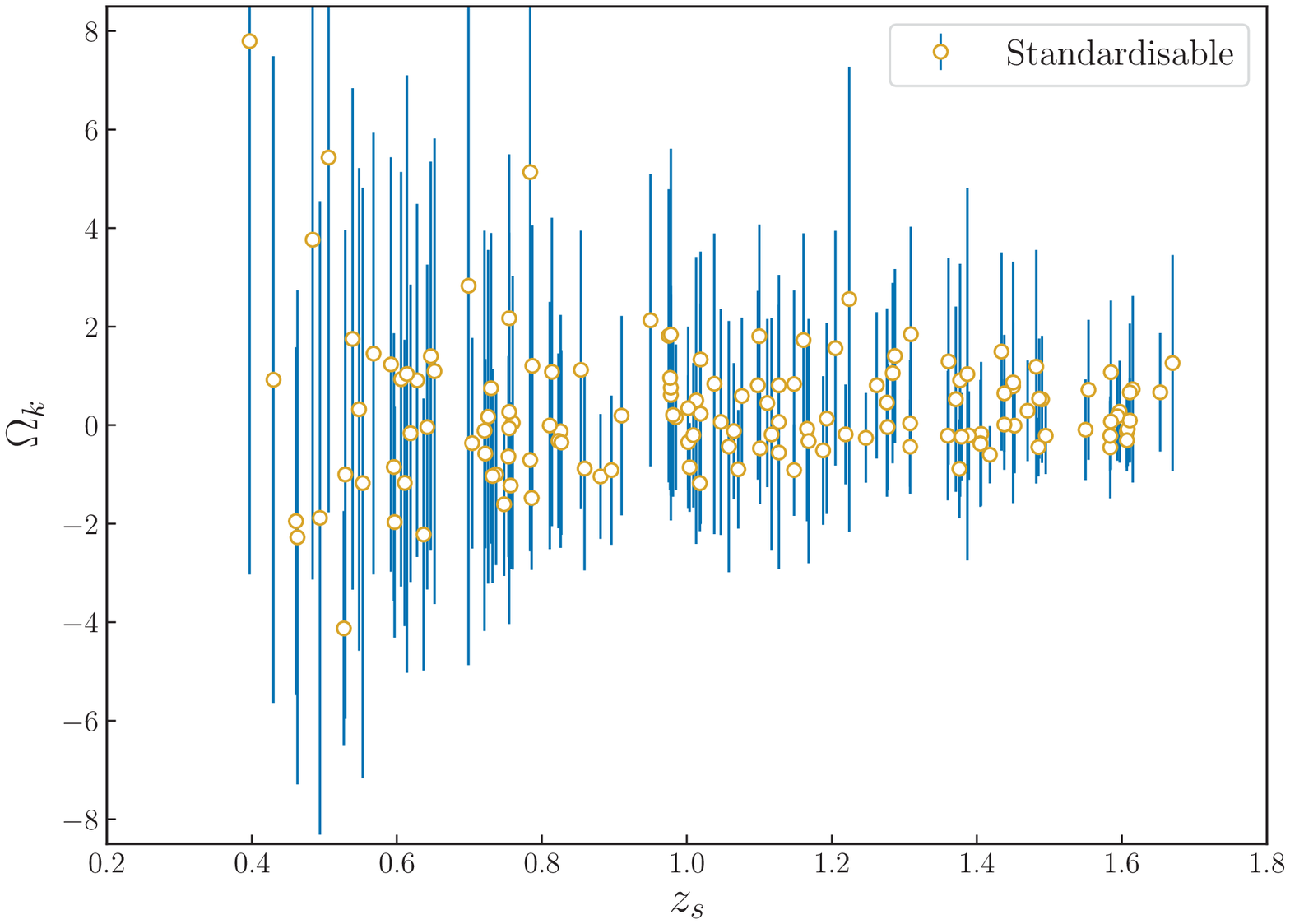} \includegraphics[scale=0.3]{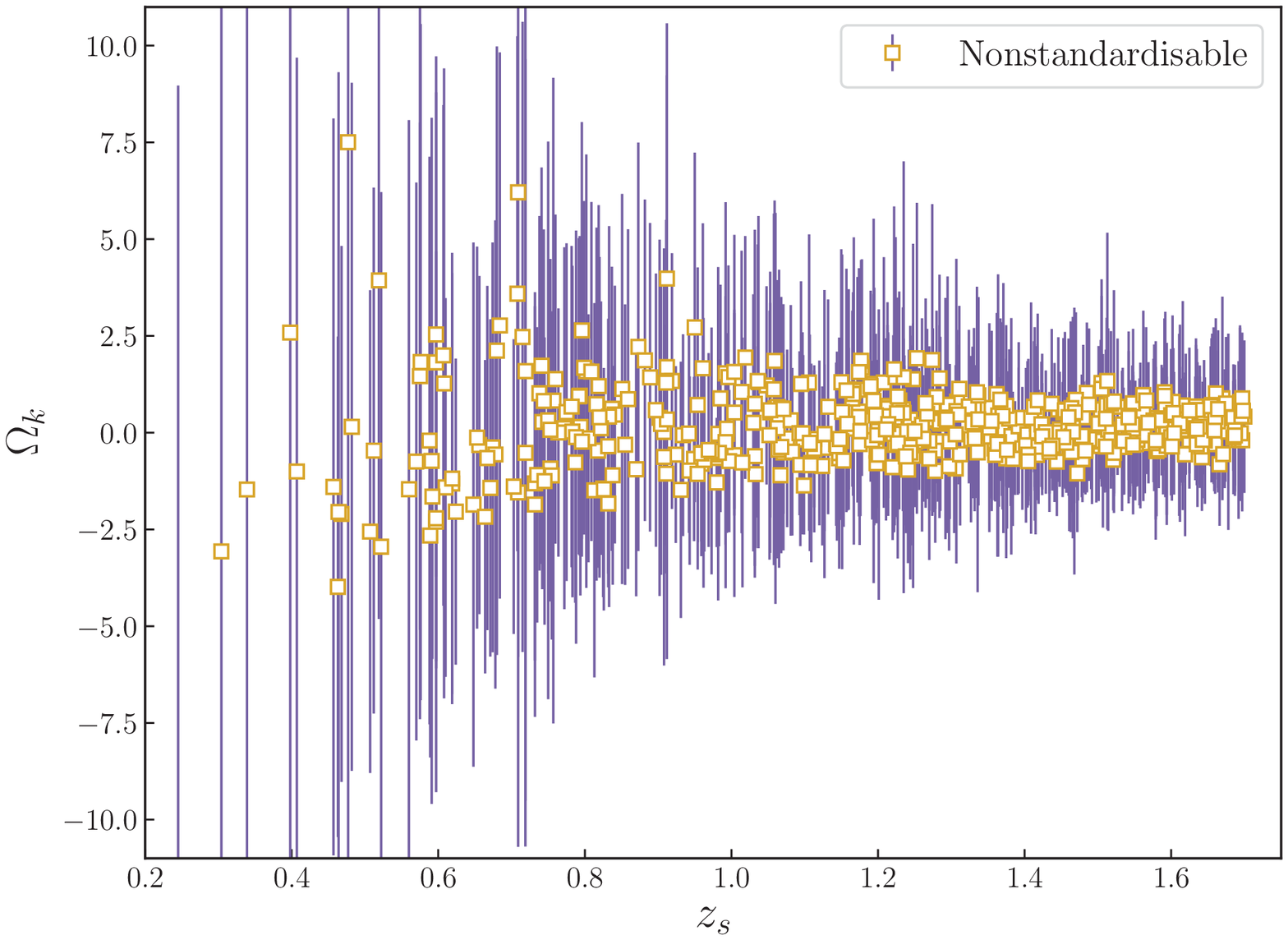}
\caption{ An example of the simulated measurements of $\Omega_k$
from future observations of SGLSNe Ia: without and with the effect
of microlensing. \textbf{The blue lines denote the associated error
bars (68.3\% C.L.) of $\Omega_k$ when all the uncertainties are
included.} }\label{fig1}
\end{figure}

\begin{table}
\begin{center}
{\scriptsize%\footnotesize %\tiny
\begin{tabular}{lcccc}
\hline\hline

& $\delta\theta_E$ & $\delta\sigma_{ap}$ & $\delta \gamma$  \\
\hline
Multiple images  &1\% & 5\% & 1\%\\
\hline
& $\delta\Delta t$  & $\delta\Delta t$ (ML) & $\delta\Delta\psi$   &  $\delta\Delta\psi$ (LOS) \\
\hline
Time delay  &1\%  & 1\% & $\propto(\delta\theta_E,\delta \gamma)$   &1\%\\
\hline & $\Delta \mu_D (sta)$ & $\Delta\mu_D$ (ML)   & $\delta \mu$ & $\Delta \mu_D (sys)$  \\
\hline
Lensed SNe Ia & $\sigma_{stat}$ & 0.70 mag & $\propto(\delta\theta_E,\delta \gamma)$ & $\sigma_{sys}$    \\
\hline\hline
\end{tabular} }
\end{center}
\caption{ Uncertainties of contributing to the uncertainty of
$\Omega_k(z_l,z_s)$ measurement (ML and LOS correspond to the
microlensing effect and light-of-sight contamination, respectively).
Concerning the Fermat potential and SNe Ia images magnifications by
the lensing galaxy potential, relevant uncertainties were simulated
by propagating uncertainties of $\theta_E$ and
$\gamma$.}\label{error}
\end{table}

\textit{Simulated data.---} Recent analysis \citep{Goldstein17}
revealed that the LSST can discover up to 650 multiply imaged SNe Ia
in a 10 year $z$-band search. Following \citet{Collett15}, we
simulated a realistic population of SGLSNe Ia lensed by early-type
galaxies, assuming distributions of velocity dispersions and
Einstein radii similar to the SL2S sample \cite{Sonnenfeld13a}. The
velocity dispersion function of the lenses in the local Universe
follows the modified Schechter function \cite{Choi07,Cao12a}. The
population of strong lenses is dominated by galaxies with velocity
dispersion of $\sigma_{ap}=210\pm50$ km/s, while the lens redshift
distribution is well approximated by a Gaussian with mean 0.40.
Although discovering strong lenses in future surveys will require
the development of new methods and algorithms, we are confident that
the simulated population of lenses is a good representation of what
the future LSST survey might yield \cite{Collett15}. In our fiducial
model, the average logaritmic density slope is modeled as
$\gamma=2.09$ with 10\% intrinsic scatter, the results from SLACS
strong-lens early-type galaxies with direct total-mass and
stellar-velocity dispersion measurements \cite{Koopmans09}. Then we
performed a Monte Carlo simulation to create the lensed SNe Ia
sample. In each simulation, there were 650 type Ia supernova
covering the redshift range of $0.00<z \leq 1.70$. When calculating
the sampling distribution (number density) of the SNe Ia population,
we adopted the redshift-dependent SNe Ia rate from
\citet{Sullivan00}. More specifically, in our model of the SNe Ia
population, we took the redshift distribution of multiply imaged SNe
Ia detectable in a 10 year LSST z-band search \cite{Goldstein17},
which furthermore constituted the differential rates of lensed SNe
Ia events as a function of $z_s$. For each lensed SNe Ia, following
the suggestion of \citet{Goldstein17}, the peak rest-frame $M_B$ was
assumed to be -19.3, while the cross-filter $K$-corrections were
computed from the one-component SNe Ia spectral template
\cite{Nugent02}.

I. For a specific SGL system observed with HST-like image quality,
the state-of-the-art lens modeling techniques \citep{Suyu10,Suyu12b}
and kinematic modeling methods \citep{Auger10,Sonnenfeld12} enable
high precision inference of $M_E$, which can be measured to within
1-2\% including all random and systematic uncertainties
\citep{Treu10}. Following the analysis of \citet{Collett16},
fractional uncertainties of the observed velocity dispersion and the
Einstein radius are 5\% and 1\%, respectively. Although the
line-of-sight contamination might introduce 3\% uncertainties
\citep{Hilbert09}, this systematics might be reduced to 1\% in
future strong lensing surveys. Recent analysis of the SL2S lens
sample demonstrated that the total mass-density slope $\gamma$
inside the Einstein radius can be determined with 5\% accuracy
\citep{Ruff11,Sonnenfeld13b}. However, the inclusion of time delays
can reduce this uncertainty to 1\% \citep{Wucknitz04}.

II. Four sources of uncertainty are included in our simulation of
time-delay measurements: $\Delta t$ measurement itself, the effect
of microlensing, uncertainty of Fermat potential determination and
line of sight contamination. SNe Ia have many advantages over AGNs
and quasars \citep{Goldstein17}, where the new curve shifting
algorithms \citep{Tewes13a} enable $\Delta t$ measurements with 3\%
accuracy \citep{Fassnacht02,Tewes13b,Liao2015}. Time delays measured
with lensed SNe Ia are supposed to be very accurate due to the
exceptionally well-characterized spectral sequences and relatively
small variation in quickly evolving light curve shapes and color
\citep{Nugent02,Pereira13}. We assumed $\Delta t$ uncertainty  of
1\%, which seems reasonable. Next, the microlensing generated by
stars in lensing galaxy, may significantly magnify lensed supernovae
\citep{Dobler06,Bagherpour06}. Concerning LSST, the distribution of
absolute time delay error due to microlensing is unbiased at the
sub-percent level with color curve observations in the achromatic
phase \cite{Goldstein18}. Therefore, an additional 1\% uncertainty
of $\Delta t$ is added for SGLSNe Ia in which the microlensing is
significant. The uncertainty of the Fermat potential difference is
simulated from the lens mass profile and the Einstein radius
uncertainties \citep{Suyu13}. In a system with the lensed SN Ia
image quality typical to the HST observations $\sim 3\%$ precision
on the Fermat potential difference \citep{HOLI} can be achieved.
Finally, another 1\% uncertainty of $\Delta\psi$ will be considered
due to LOS effects \citep{Liao2017}.

III. Three sources of uncertainties are included in our simulation
of SGLSNe Ia. Following the strategy of \citep{Spergel15}, the
distance precision per SNe is $\sigma^2_{\rm stat}= \sigma^{2}_{\rm
meas}+ \sigma^{2}_{\rm int} + \sigma^{2}_{\rm lens}$
\citep{Hounsell17}, with the mean uncertainty $\sigma_{\rm
meas}=0.08$ mag, the intrinsic scatter uncertainty $\sigma_{\rm int}
= 0.09$~mag, and the lensing uncertainty due to the LOS mass
distribution $\sigma_{\rm lens} = 0.07 \times z$~mag
\citep{Holz05,Jonsson10}. Moreover, the total systematic uncertainty
is modeled as $\sigma_{sys} = 0.01(1 + z)/1.8~\rm{(mag)}$, which is
assumed to increase with redshift \citep{Hounsell17}. It should be
stressed that the derivation of such systematic uncertainty is based
on an  uncorrelated SNe Ia sample, while there are known systematics
contradicting this assumption (e.g., uncertainties related to
calibration and SNe Ia color are correlated across a wide redshift
range). The correlation between different SNe Ia might constitute an
important systematic error in our $\Omega_k$ measurements. The
statistical and systematic uncertainty are combined to produce the
total uncertainty as $\sigma^2_{tot} = \sigma^2_{stat} +
\sigma^2_{sys}$. Being standardizable candles, SGLSNe Ia can be used
to assess the lensing magnification factor $\mu$ directly
\citep{Oguri03} by solving the lens equation using \emph{glafic}
\citep{Oguri10b}. We explicitly considered that uncertainty of $\mu$
is related to uncertainties of $\gamma$ and $\theta_E$
\citep{Suyu13}. Finally, only 22\% of the 650 SGLSNe Ia discovered
by LSST will be standardisable, due to the microlensing effect
\citep{Foxley-Marrable18}. Lensed images are standardisable in
regions of low convergence, shear and stellar density (especially
the outer image of an asymmetric double for lenses with large
$\theta_E$). Therefore an additional uncertainty $\sim 0.70$ mag
should be considered for the remaining 78\% of SGLSNe Ia
\cite{Yahalomi17}, especially in quadruple image systems, symmetric
doubles and small Einstein radii lenses. All images are used to
determine $\mu$ and SNe Ia distances. Table I lists the relative
uncertainties of factors contributing to the accuracy of
$\Omega_k(z_l, z_s)$ measurements.

We summarize the main route of our method as follows. There are
three levels of random realizations that need to be simulated
separately: 1) Monte Carlo simulation of the strong lensing systems;
2) the statistical errors that are independent for each lensing
system; and 3) the systematic errors whose realizations are common
for all systems. Specifically, in this analysis, the total
systematic uncertainty is considered in the corrected SNe Ia
distances, which is modeled as $\mu_D (sys) = 0.01(1 +
z)/1.8~\rm{(mag)}$ \citep{Hounsell17}. With the combination of these
different layers of randomness, we use Monte-Carlo simulation to
project the statistical uncertainties of the observables of
$\gamma$, $\theta_E$, $\theta_{ap}$, $\Delta t$, and $\mu_D$, as
well as the systematic uncertainty of the SNe Ia distances ($\mu_D$)
onto the final uncertainty of $\Omega_k$. In order to guarantee the
reliability of our results, we realize $10^3$ random mock data sets
and apply the above algorithm to each of them.

\begin{figure}
\includegraphics[scale=0.33]{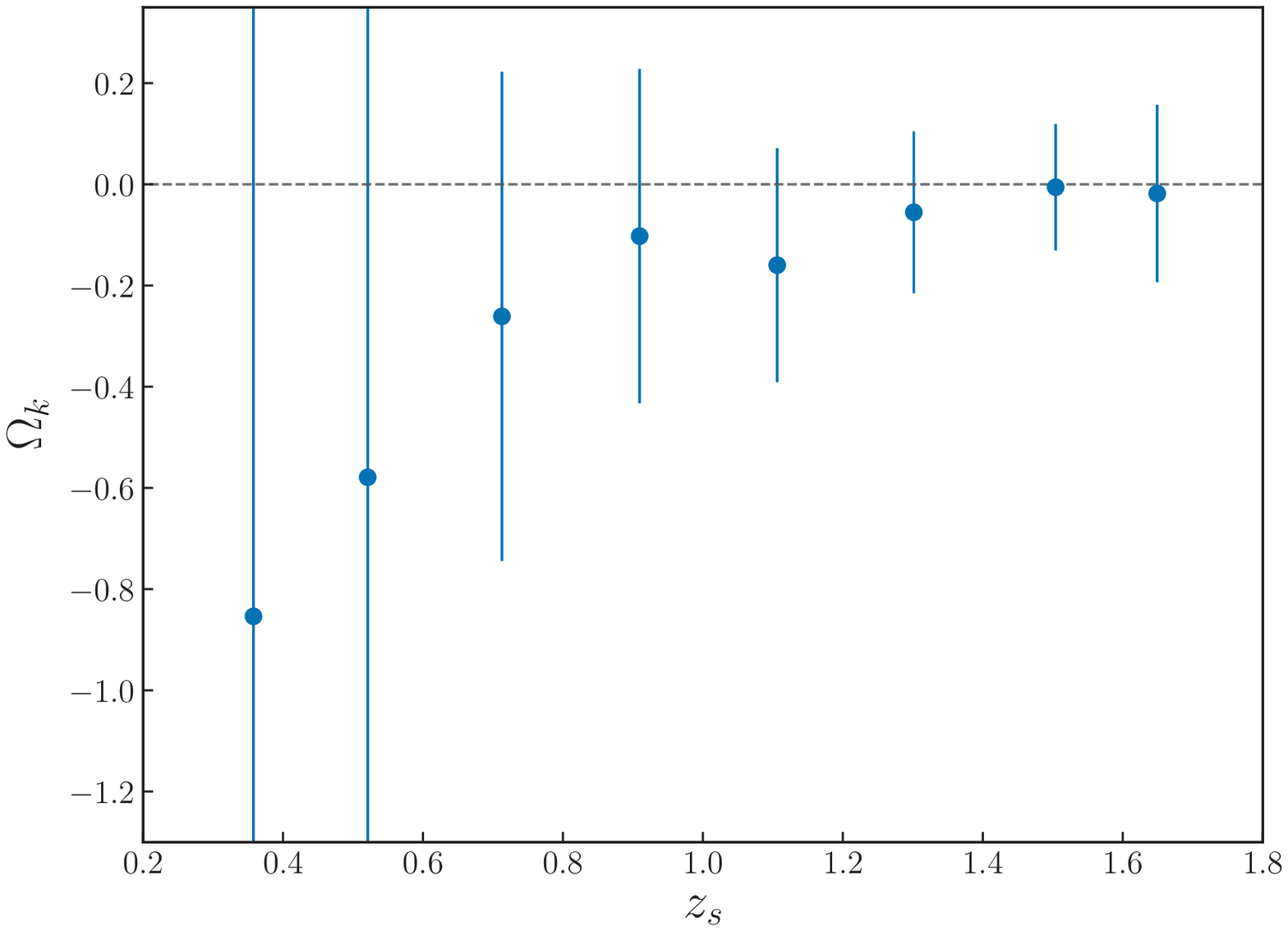}
\includegraphics[scale=0.3]{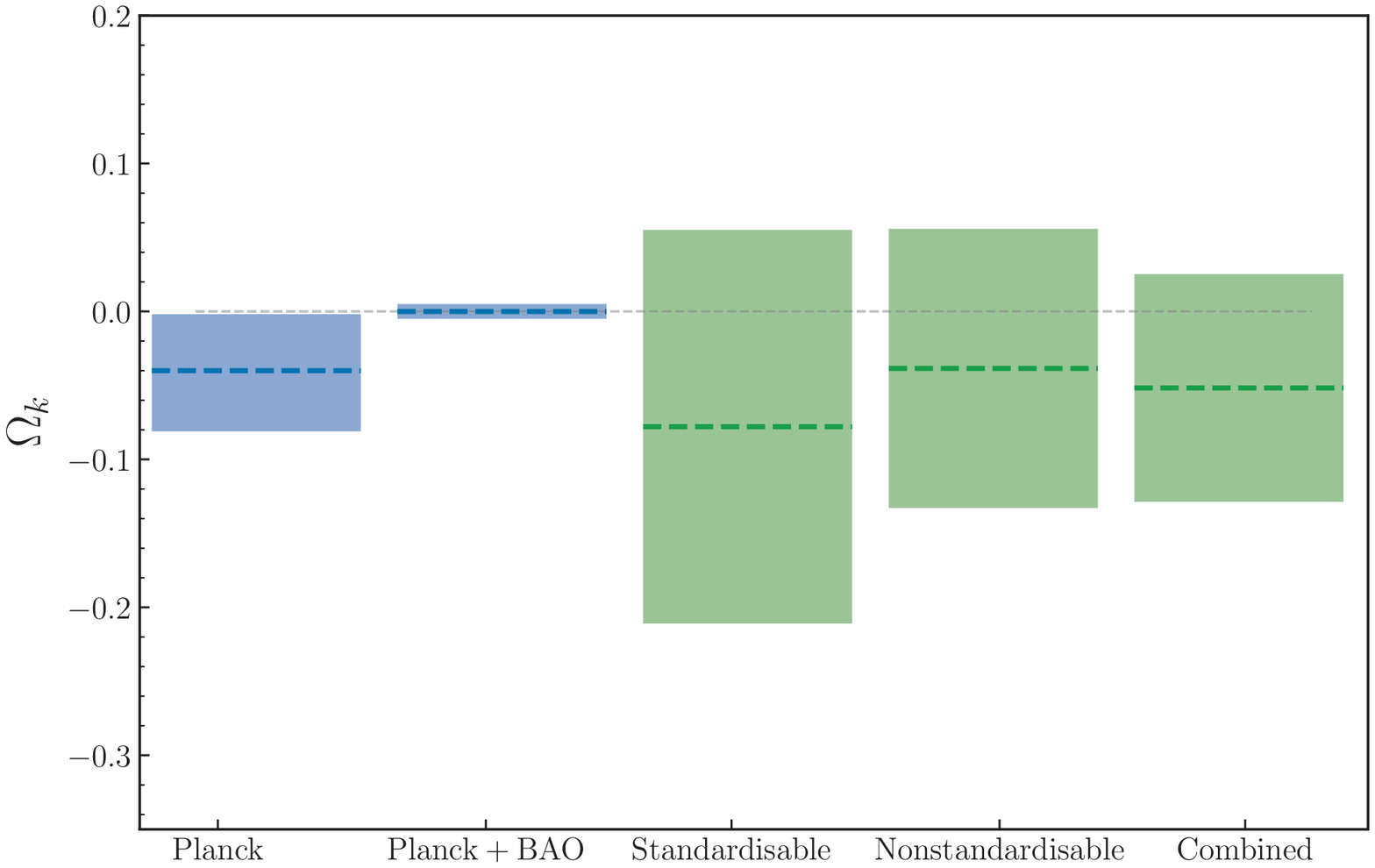}
\caption{ Upper panel: simulated measurements of $\Omega_k$ divided
into $z_s$ bins and summarized as weighted means with corresponding
standard deviations. Lower panel: inverse variance weighted mean of
$\Omega_k$ achievable from SGLSNe Ia in comparison to CMB+BAO
constraints. }\label{fig2}
\end{figure}

\textit{Constraint results.---} Assuming that parameters whose
uncertainties listed in Table I follow the Gaussian distribution, we
simulated two sets of realistic lensed SNe Ia with and without
considering the effect of microlensing. Concerning the error budget
applied in this paper, one should clarify that the objective of the
work is to determine uncertainties in $\Omega_k$ for a survey.
However, in each of the $10^3$ Monte Carlo simulations, new surveys
with independent sets of SGLSNe Ia systems are realized, which
indicates that the uncertainties inherent in having one survey with
one realization of SGLSNe Ia systems are underestimated. Moreover,
since all of these SNe Ia are strongly lensed, the lensing
dispersion of $\mu_D$ is also correlated with the other parameters.
Therefore, in this analysis, we combined the error budget assuming
zero measurement uncertainty and that assuming zero per-object
intrinsic dispersion (the intrinsic magnitudes dispersion of
$\mu_D$). An example of the simulation is shown in Fig.~1 (based on
one realization of SGLSNe Ia sample with one realization of
statistical and systematic errors), which were repeated $10^3$ times
to produce the statistical results shown in Figs.~2-3 (based on
$10^3$ realizations of SGLSNe Ia sample with $10^3$ realizations of
statistical and systematic errors). Turning to the mock SGLSNe Ia
catalogue of \citep{Goldstein17}, only 22\% of the full sample
discovered by LSST will be standardisable. Such conclusion is
consistent with the predicted relation between the standardisable
fraction and Einstein radius assuming the Salpeter IMF
\citep{Foxley-Marrable18}, which implies that 90\% of the source
plane with $\theta_E \geq 1"$ is standardisable. In our simulated
data, the mean Einstein radius for standardisable SGLSNe Ia is 1.5",
compared to 0.73" for SGLSNe Ia unsuitable to be standard candles.
The mean time delay for standardisable LSST SGLSNe Ia is 71 days,
compared to 36 days for non-standardisable ones. \emph{Are these
measurements sufficient enough to detect possible deviation from
FLRW metric?} As can be seen in Fig.~1, relatively low precision of
individual $\Omega_k(z_l, z_s)$ measurements, especially in
low-redshift SNe Ia, makes it very difficult to be competitive.
However, at higher redshifts one would be able to find different
$\Omega_k(z_l, z_s)$ in different pairs of $(z_l, z_s)$, which could
indicate that light propagation on large scales was affected by
departures from the FLRW metric.

\emph{Is it possible to achieve a stringent measurement of the
spatial curvature from a statistical sample of SGLSNe Ia?} One
should be very careful to the bias induced by the $\Omega_k$
measurements with large uncertainties. The most straightforward way
of summarizing multiple measurements is inverse variance weighting
\citep{Zheng16,Denissenya18}
\begin{equation}
\begin{array}{l}
\bar{\Omega_k}=\frac{\sum\left(\Omega_{k,i}/\sigma^2_{\Omega_{k,i}}\right)}{\sum1/\sigma^2_{\Omega_{k,i}}},\\
\sigma^2_{\bar{\Omega_{k}}}=\frac{1}{\sum1/\sigma^2_{\Omega_{k,i}}},
\end{array}
\end{equation}
where $\bar{\Omega_{k}}$ stands for the weighted mean of cosmic
curvature with uncertainty $\sigma_{\bar{\Omega_{k}}}$. In order to
get a better feeling of the bias inherent to the $\Omega_k(z_l,
z_s)$ observable due to its complex nonlinear dependence on
observable quantities, we subdivided simulated 650 data points into
8 redshift bins of width $\Delta z=0.2$. The result is shown in
Fig.~2 where weighted means and corresponding standard deviations
are shown for each bin, allowing a direct check of its predicted
constancy with redshift. It is worth noticing that, although there
is a bias in the weighted mean $\Omega_k$ with large uncertainties
at low-redshifts, the mean value of the cosmic curvature is still
located within the error bar (68.3\% C. L.). A thorough discussion
of such biases and a proposal for remedy was given in
\citet{Denissenya18}. However, the observational setting discussed
in this interesting and important paper was different from ours.
Second panel of Fig.~2 shows the constraints (inverse variance
weighted mean) achievable from the full SGLSNe Ia in comparison to
CMB+BAO model-dependent constraints.

Using only standardizable SNe Ia we are able to constrain the cosmic
curvature parameter with the precision of $\Delta\Omega_k=0.13$. The
remaining $78\%$ corrected for the microlensing effect, give
$\Delta\Omega_k=0.09$. Finally, the full sample of 650 lensed SNe Ia
will improve the constraint to $\Delta\Omega_k=0.08$. Our method
might perform better, with the Chabrier IMF more lensed SNe Ia can
be classified as standard candles: lenses with smaller Einstein
radius ($\theta_E \sim 0.4"$) can have a source plane which is 90\%
standardisable \citep{Foxley-Marrable18}. Namely, with 650 simulated
SGLSNe Ia, the cosmic curvature parameter can be determined to
$\Delta \Omega_k=0.04$, which is comparable to that of the power
spectra (TT,TE,EE+lowP) from the \textit{Planck} 2015 results
\citep{Ade16}. Therefore, by comparing spatial curvature from SGLSNe
Ia with the constraints obtained from CMB, one will be able to test
the validity of the FLRW metric. The curvature determined from CMB
and BAO is impressively consistent with flat universe $\Delta
\Omega_k=0.005$ \citep{Ade16} and future CMB missions and BAO
surveys (CORE + DESI) are expected to constrain curvature at $\Delta
\Omega_k=0.0008$ \cite{CORE}. We emphasize an alternative nature of
our method. Stringent constraints could be obtained through the
CMB+BAO combined analysis (see Fig.~2) in a model dependent way.
However, assessment of the spatial curvature based on local objects
like strongly lensed supernovae would be of paramount importance
being able to probe deviations from the FLRW metric caused by the
structure formation, which is inaccessible to the tools like CMB,
BAO, and the interpolations of redshift differences detected from
galaxy redshift surveys \cite{Alam15,Laureijs11}.

\begin{figure}
\includegraphics[scale=0.3]{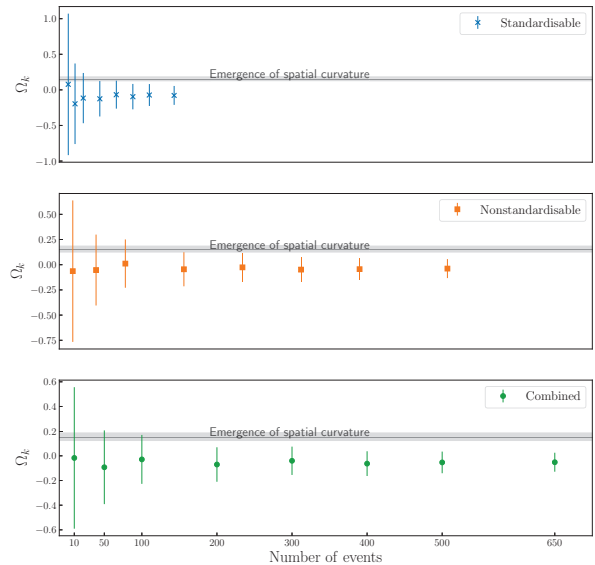}
\caption{ Inferred $\Omega_k$ parameter as a function of the number
of SGLSNe Ia, with the prediction of a silent universe added for
comparison.}\label{fig3}
\end{figure}

\emph{Is it possible to confirm or falsify alternative approaches
like the back-reaction from inhomogeneities?}
\citet{Bolejko17a,Bolejko17b} examined the so-called silent universe
in which the back-reaction of inhomogeneities was taken into
account. The most striking conclusion of these works is the
emergence of spatial curvature in the low redshift universe:
$\Omega_k=0.15^{+0.04}_{-0.03}$ (95\% confidence level). Note
studies of the local universe encompass a region with redshifts
approximately less than 0.1, where there is a linear Hubble Flow and
low redshift-data (SNe Ia) can be observed in most detail. It is
interesting to see if our method can be used to test this
prediction. Fig.~3 shows the precision of the curvature parameter
assessment as a function of SGLSNe Ia sample size. One can see that,
even with 50 SGLSNe Ia one can effectively differentiate between the
silent universe and the concordance $\Lambda$CDM cosmology, which
means that the phenomenon of emerging curvature will soon be
directly testable with observational data and furthermore
strengthens the probative power of our method to inspire new
observing programs or theoretical work in the moderate future.

In order to implement our method, dedicated observations including
spectroscopic redshift measurements of the lens and the source,
velocity dispersion of the lens, higher angular resolution imaging
to measure the Einstein radius, and dedicated campaigns to measure
time delays would be necessary. Obtaining these data for a sample of
several hundreds of SGLSNe Ia would require substantial follow-up
efforts, similar to that made in strong gravitational lens ESO
325-G004 \citep{Collett18}. Despite of these difficulties, one may
expect that multiple measurements of SGLSNe Ia can become an
independent alternative to current probes
\cite{Cao17a,Cao17b,Ma17,Qi17,Xu18,Cao19}, useful for more precise
empirical studies of the FLRW metric. Finally, concerning the
probative power of our method, the methodology proposed in this
paper might be extended to strongly lensed gravitational wave (GW)
events detected by aLIGO and the proposed ET \citep{Li18}. Given the
wealth of available gravitational lensing data in EM and GW domain,
we may be optimistic about detecting possible deviation from the
FLRW metric within our observational volume in the future. Such
accurate model-independent measurements of the FLRW metric can
become a milestone in precision cosmology.

\textit{Acknowledgements.---} We are grateful to Paul L. Schechter
for useful discussions. This work was supported by National Key R\&D
Program of China No. 2017YFA0402600, the National Natural Science
Foundation of China under Grants Nos. 11690023, 11373014, and
11633001, Beijing Talents Fund of Organization Department of Beijing
Municipal Committee of the CPC, the Strategic Priority Research
Program of the Chinese Academy of Sciences, Grant No. XDB23000000,
the Interdiscipline Research Funds of Beijing Normal University, and
the Opening Project of Key Laboratory of Computational Astrophysics,
National Astronomical Observatories, Chinese Academy of Sciences.
J.-Z. Qi was supported by China Postdoctoral Science Foundation
under Grant No. 2017M620661, and the Fundamental Research Funds for
the Central Universities N180503014. This research was also partly
supported by the Poland- China Scientific \& Technological
Cooperation Committee Project No. 35-4. M. Biesiada was supported by
Foreign Talent Introducing Project and Special Fund Support of
Foreign Knowledge Introducing Project in China.

\end{document}